%% file: Submit3.tex
\documentclass[12pt]{article}
\usepackage{graphicx}

\usepackage{hyperref}
\usepackage{cite}

\usepackage{amssymb,amsmath,latexsym,graphics, graphicx,epsfig} 
\usepackage{latexsym}
\usepackage{amssymb}
\usepackage{epsfig,amsmath,graphics}
\usepackage{multirow}
\usepackage{appendix}
\usepackage{slashed}
\usepackage{anysize}

\newcommand{\gsim}{\lower.7ex\hbox{$\;\stackrel{\textstyle>}{\sim}\;$}}
\newcommand{\lsim}{\lower.7ex\hbox{$\;\stackrel{\textstyle<}{\sim}\;$}}

\def\OO{{\cal O}}

\newcommand{\bef}{\begin{figure}[htbp]\begin{center}}
\newcommand{\eef}{\end{center}\end{figure}}

\newcommand{\bea}{\begin{eqnarray}}
\newcommand{\eea}{\end{eqnarray}}

  \def\doeack{\footnote{Work supported by the US Department of Energy,
                     contract DE--AC02--76SF00515.}}

\newcommand\pubnumber{SLAC--PUB--14457}
\newcommand\pubdate{May, 2011}

\def\SLAC{SLAC,
    Stanford University, Menlo Park, California 94025 USA}

\def\Title#1{\begin{center} {\Large #1 } \end{center}}
\def\Author#1{\begin{center}{ \sc #1} \end{center}}
\def\Address#1{\begin{center}{ \it #1} \end{center}}

\newcommand\pubblock{\rightline{\begin{tabular}{l} \pubnumber\\
         \pubdate \end{tabular}}}
\newenvironment{Abstract}{\begin{quotation} \begin{center}
                       ABSTRACT
     \end{center}\bigskip  }{\end{quotation}}

\input mymacros.tex

\begin{document}
\begin{titlepage}
\pubblock

\Title{Unitarity constraints on asymmetric freeze-in}
\Author{ Anson Hook\doeack}
\Address{\SLAC}
\begin{Abstract}

This paper considers unitarity and CPT constraints on asymmetric freeze-in, the use of freeze-in to store baryon number in a dark sector.  In this scenario, Sakharov's out of equilibrium condition is satisfied by placing the visible and hidden sectors at different temperatures while a net visible baryon number is produced by storing negative baryon number in a dark sector.  It is shown that unitarity and CPT lead to unexpected cancellations.  In particular, the transfer of baryon number cancels completely at leading order.


\end{Abstract}
\end{titlepage}
\def\thefootnote{\fnsymbol{footnote}}
\setcounter{footnote}{0}

\tableofcontents

\section{Introduction}


Models of baryogenesis require CP violation, baryon number violation and a process that goes
out of thermodynamic equilibrium~\cite{Sakharov:1967dj}¿.
  For processes in thermal equilibrium,
the constraints on the S-matrix that come from unitarity are already
strong enough to prohibit the production of a net baryon number.
In most models of baryogenesis, the out of equilibrium situation is 
produced by `freeze-out'; a heavy particle is sufficiently
long-lived that it survives into an era when it is not being produced 
by thermal processes.  In the presence of B and CP violation, the decays of 
this particle can be asymmetric, producing nonzero baryon 
number \cite{Kuzmin:1985mm,Nanopoulos:1979gx,Cline:2006ts,Riotto:1998bt,Chen:2007fv,Giudice:2003jh,Davidson:2008bu,Luty:1992un}.

Recently, \cite{Hall:2009bx} suggested another 
mechanism for obtaining an out-of-equilibrium situation.  Their
mechanism, called `freeze-in', involves a `dark' sector
that never achieves thermal equilibrium with the `visible' sector, 
 particles with Standard Model quantum numbers.  This 
hidden sector interacts with the visible sector through a small coupling.  The products of these 
reactions reflect the out-of-equilibrium nature of the dark sector.
In \cite{Hall:2009bx}, freeze-in is used to produce dark matter.  \cite{Hall:2010jx} proposed using freeze-in as a mechanism for transferring baryon number to a dark sector, creating an apparent baryon asymmetry; this mechanism was termed asymmetric freeze-in.  Asymmetric freeze-in is similar in spirit to Dirac leptogenesis with a different production mechanism \cite{Dick:1999je,Murayama:2002je,Thomas:2006gr}.  The idea of using different temperature sectors when generating baryon abundances in the visible and dark sectors was also used in the mirror world models \cite{Berezhiani:2000gw,Bento:2001nb,Bento:2002sj,Ciarcelluti:2004ik,Ciarcelluti:2004ip,Foot:2003jt,Foot:2004pq,Berezhiani:2003wj,Ciarcelluti:2010zz,Foot:2010hu}.

Motivated by the asymmetric freeze-in scenario, this paper considers the transfer of baryon number or any other $U(1)$ charge between two sectors which are at different temperatures.  
The unitarity of the S-matrix places strong constraints on the transfer of a $U(1)$ charge and forces the
leading contribution to cancel.  Specifically, let the coupling between the two
sectors be given by a small parameter $\lambda$.  The $U(1)$ charge transfer between the two sectors cancels at order $\lambda^2$, so that net baryon
number is generated only at order $\lambda^3$.   This result has an important effect on the analysis in \cite{Hall:2010jx}.
It dramatically reduces the region of parameter space in which the mechanism of \cite{Hall:2010jx}
is effective.  However, beyond this application, the cancellation is an interesting and nontrivial property of thermal field theory that might well have other applications. 

This paper is organized as follows: Sec.~\ref{Sec: Notation} introduces some 
useful notation.  Sec.~\ref{Sec: Example} gives an explicit example that shows how baryon number generation cancels to leading order in asymmetric freeze-in. Sec.~\ref{Sec: Unitarity} gives a general proof that unitarity requires the cancellation of baryon number generation at order $\lambda^2$, analogous to the standard unitarity argument preventing baryon number violation at leading order \cite{Nanopoulos:1979gx}.
Sec.~\ref{Sec: Diagrams} shows how the cancellation of baryon number production
at order $\lambda^2$ appears from Feynman diagrams and exhibits effects that generate baryon number at $\lambda^3$.  
In Sec.~\ref{Sec: Deviations}, deviations from thermal equilibrium are briefly discussed.

\section{Notation}
\label{Sec: Notation}

To reduce clutter, a few shorthand notations will be used.  CP violation involves differences between particle and antiparticle production.  Thus, let
\bea
\label{Eq: Rate}
 |\mathcal{M}(\alpha \rightarrow \beta)|^2  - |\mathcal{M}(\bar{\alpha} \rightarrow \bar{\beta})|^2  \equiv \Delta |\mathcal{M}(\alpha \rightarrow \beta)|^2
\eea
 The terminology of visible and dark sectors and baryon number are only meant to be suggestive.  The results of this paper hold for the transfer of any $U(1)$ charge, $Q$, between two weakly coupled sectors.   Each sector is in thermal equilibrium with itself, but the two sectors have different temperatures.

A state is represented by a Greek letter while visible and hidden sector states are distinguished by their subscripts.  A general state is simply the tensor product $\alpha = \alpha_v \alpha_h$ where either of the states $\alpha_v$ or $\alpha_h$ may be the vacuum state, $\Omega$.

A product of phase space densities, $f$, appear in the Boltzmann equations.
The phase space densities of the initial state are combined into the notation
\bea
\label{Eq: phase}
f_{\alpha} &=& \prod_i f_{{\alpha(i)}} 
\eea
The Pauli exclusion/stimulated emission factors of the final state will be written as
\bea
\label{Eq: phase2}
(1 \pm f_{\alpha}) &=& \prod_i (1 \pm f_{\alpha(i)})
\eea
 In both formulas, the product runs over all the particles in a state.  The two sectors are assumed to be in thermal equilibrium with themselves so that if there exists a process $\mathcal{M}(\alpha_v \Omega_h \rightarrow \beta_v \Omega_h)$, $f_{\alpha_v} (1 \pm f_{\beta_v}) = f_{\beta_v} (1 \pm f_{\alpha_v})$.  This is consistent with thermal equilibrium through the identity 
 \bea
 \label{Eq: expo}
 \frac{f_{\alpha}} {1 \pm f_{\alpha}} = \prod_i e^{-\frac{E_{\alpha(i)}-\mu_{\alpha(i)}}{T}}
 \eea
since the sums of the particle energies and chemical potentials balance in equilibrium.  This identity is a direct consequence of Boltzmann's H theorem \cite{Kolb:1979qa}.

\section{Example}
\label{Sec: Example}

This section gives an explicit example of the cancellation of $U(1)$ charge transfer at $\OO(\lambda^2)$.  Sec.~\ref{Sec: sub} gives some background and an explicit calculation is done in Sec.~\ref{Sec: Model}.

\subsection{Background}
\label{Sec: sub}

In baryogenesis, the matrix elements that contribute to the Boltzmann equation for particle densities are computed with finite temperature Feynman diagrams.  Thermal field theory propagators contain delta function interactions with the thermal bath.  The cancellation of CP asymmetries at leading order relies critically on this modification of the propagator.  

In zero temperature field theory, the familiar matrix elements take into account all possible ways of going from the initial state to the final state.  However, the Boltzmann equations already take into account the classical evolution of states.  Including on-shell processes which go from state $\alpha$ to $\beta$ to $\gamma$ both in the matrix elements and in the Boltzmann equation would be double counting.  
Double counting is avoided by removing all matrix elements which can be interpreted as a classical evolution from state $\alpha$ to $\beta$ to $\gamma$.  This is a well-known procedure called real intermediate state subtraction (RISS) \cite{Kolb:1979qa}.  If there is a matrix element $\mathcal{M}(\alpha \rightarrow \beta \rightarrow \gamma)$ where $\beta$ goes on-shell, the subtraction scheme removes the on-shell contribution 
\bea
|\mathcal{M}_{\text{RISS}}(\alpha \rightarrow \gamma)|^2 = |\mathcal{M}(\alpha \rightarrow \gamma)|^2 - |\mathcal{M}_{\text{NWA}}(\alpha \rightarrow \beta \rightarrow \gamma)|^2
\eea
where NWA stands for the narrow width approximation.  Roughly speaking, we remove any matrix element which requires the use of the narrow width approximation.
\begin{figure}[tbh] 
   \centering
   \includegraphics[width=6in]{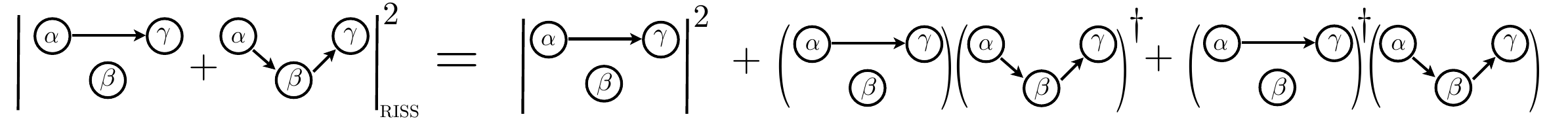} 
   \caption{The Boltzmann equations account for the classical evolution of states.  Real intermediate state subtraction avoids double counting by removing matrix elements which are already taken into account by the Boltzmann equations.}
   \label{Fig: BDD}
\end{figure}

Real intermediate state subtraction greatly simplifies CP violation.  Without the subtraction, diagrams in which states go on-shell are regulated by the resummed width $\Gamma$.  Since $\Gamma$ appears in the denominator, higher-order diagrams can cancel lower order diagrams.  These cancellations must occur; they are required by the unitarity of the unsubtracted matrix elements. A simple example of this phenomenon is given in Appendix \ref{Sec: B}.

Since the width is not needed to regulate subtracted diagrams, it is valid to consider the diagrams with the widths of particles $\bf{not}$ resummed.  For every partial width, there is a different diagram that will cancel its CP asymmetry.  While real intermediate state subtraction is unimportant to the unitarity proof in Sec.~\ref{Sec: Unitarity}, in order to see the cancellation as one diagram canceling against one other diagram, the widths must not be resummed.

The matrix elements considered will describe processes $\alpha_v \alpha_h \rightarrow \beta_v \beta_h$ with $Q(\alpha_v) \ne Q(\beta_v)$.  To calculate the CP asymmetry of these processes, we take $\mathcal{M}(\alpha \rightarrow \beta) = \sum_i \lambda_i F_i$ and $\mathcal{M}(\bar{\alpha} \rightarrow \bar{\beta}) = \sum_i \lambda_i^* F_i$, where $\lambda_i$ denotes the product of coupling constants and $F_i$ denotes the kinematics of the Feynman diagram.  The rate difference in Eq.~\ref{Eq: Rate} is
\bea
\Delta |\mathcal{M}(\alpha \rightarrow \beta)|^2 &=& - 2 \sum_{i,j} \text{Im}( \lambda_i \lambda_j^*) \text{Im}(F_i F_j^*) \\ 
&=& - 4 \sum_{i,j} \text{Im}( \lambda_i \lambda_j^*) \text{Re}(F_j) \text{Im}(F_i)
\eea

In order for an imaginary part in a Feynman diagram to occur, there needs to be an intermediate state, $\delta$, which can go on-shell.  Cutkosky's rule is used to find its value.  In the presence of finite density, Cutkosky's rule is modified.  The thermal bath into which particles are created must be taken into account, so an additional multiplicative factor of $(1 \pm f_{\delta})$ appears.  Interference between the process $\mathcal{M}(\alpha \rightarrow \beta)$ and $\mathcal{M}(\alpha \rightarrow \delta \rightarrow \beta)$ gives 
\bea
\label{Eq: interference}
\Delta | \mathcal{M}(\alpha \rightarrow \beta)|^2  \propto \sum_{\delta,\text{diagrams}} \int d \Pi_\delta (1 \pm f_\delta )\text{Im}[ \lambda_{\alpha \rightarrow \beta} \lambda_{\alpha \rightarrow \delta}^* \lambda_{\delta \rightarrow \beta}^* ] F_{\alpha \rightarrow \beta} F_{\alpha \rightarrow \delta} F_{\delta \rightarrow \beta}
\eea
$\lambda$ indicates the couplings associated with the diagram, $F$ indicates the real part of Feynman diagrams and $\int d \Pi_\delta$ is a phase space integral.  The sum over diagrams is simply the sum over $i$ and $j$ from before.  While $|\mathcal{M}(\alpha \rightarrow \delta \rightarrow \beta)|^2$ is removed by real intermediate state subtraction, due to its classical interpretation, the interference between $\mathcal{M}(\alpha \rightarrow \beta)$ and $\mathcal{M}(\alpha \rightarrow \delta \rightarrow \beta)$ is a quantum mechanical effect and must be included explicitly in the Boltzmann equations.  Throughout, $\lambda$ will be used for couplings between the two sectors and $g$ for couplings within the sectors.

\subsection{Model}
\label{Sec: Model}

This section uses an explicit model to show the differences between out of equilibrium decay and freeze-in.  In out of equilibrium decay, only a single particle goes out of equilibrium.  Therefore when calculating CP violation, the only initial state that is important is the decaying particle.  On the contrary, when we have two sectors at different temperatures, there are many different initial states, all of equal importance.  The decay of a single particle and 2-to-2 scattering must be treated on the same footing.  

Consider a visible sector $A$ and a dark sector $B$ where $A$ is at a nonzero temperature and $B$ is at zero temperature.  There are particles $A_1$,$A_2$ and $A_3$ in the $A$ sector and particles $B_0$,$B_1$,$B_2$ and $B_3$ in the $B$ sector.  The subscript labels their $U(1)$ charge.  Consider the most general Lagrangian allowed by symmetries where the terms in the Lagrangian which are relevant for the following discussion are
\bea
\mathcal{L} &\supset& \lambda_1 A_3 B_1^\dagger B_2^\dagger + \lambda_2 A_3 B_3^\dagger B_0^\dagger + \lambda_3 A_1 A_2 B_1^\dagger B_2^\dagger + \lambda_4 A_1 A_2 B_0^\dagger B_3^\dagger \CR
 &+& g_1 A_3 A_1^\dagger A_2^\dagger + g_2 B_3 B_0 B_1^\dagger B_2^\dagger 
\eea
All interactions allowed by the symmetries of the problem will be generated, but only the above ones will be relevant for the following discussion.  There is initially no net baryon number in either sector.  Interactions between the visible sector and the dark sector store baryon number in the dark sector, creating an apparent asymmetry.

There are two main types of cancellations in this model.  The first is illustrated by the decay of a particle $A_3$.  In a CP violating theory, $A_3$ and $\bar{A}_3$ can decay differently and generate a transfer of baryon number.  However, the total decay widths of a particle and its anti-particle are the same.  Thus, if all of the channels are weighted equally, the effect sums to zero.  The dotted lines denote Cutkosky cuts used to find the imaginary part of diagrams.
  Fig. \ref{Fig: BDD} shows the mechanism of the cancellation.  The interference of the two diagrams on the top cancels with the interference of the two diagrams on the bottom. This cancellation of CP asymmetries is well known in models of freeze-out baryogenesis \cite{Cline:2006ts}.  

\begin{figure}[tbh] 
   \centering
   \includegraphics[width=4in]{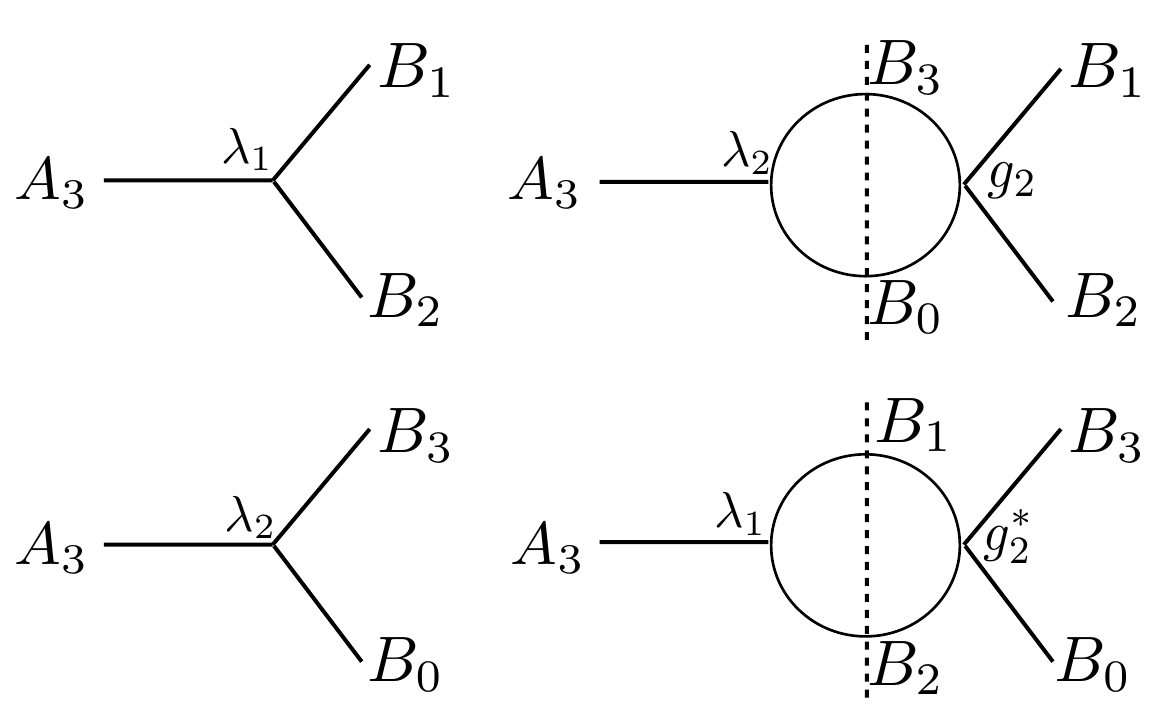} 
   \caption{These two diagrams cancel so that no net baryon number is generated in the $B$ sector.}
   \label{Fig: BDD}
\end{figure}

\begin{figure}[h] 
   \centering
   \includegraphics[width=4in]{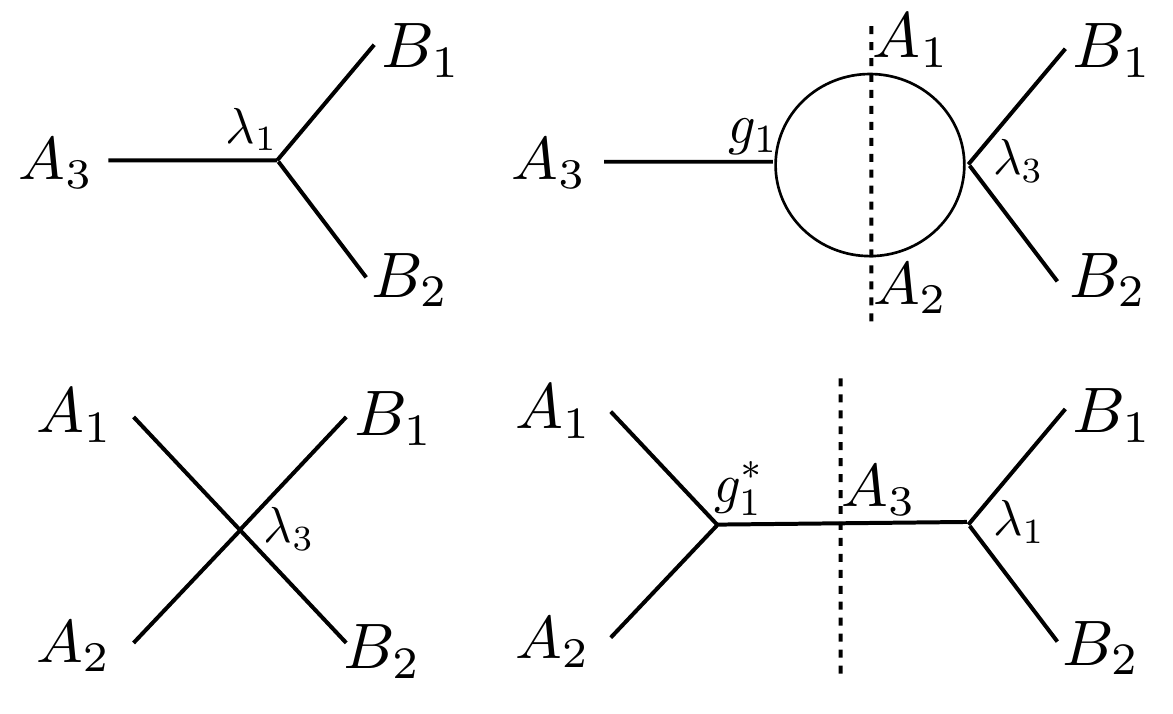} 
   \caption{Freeze-in allows for new cancellations not seen in freeze-out and decay baryogenesis.  These two diagrams show that different initial states can have canceling CP asymmetries.}
   \label{Fig: BBD}
\end{figure}
An additional type of cancellation can occur in asymmetric freeze-in as is shown in Fig. \ref{Fig: BBD}.  The top two diagrams have an interference that is not cancelled over the sum of final states because $A_3 \rightarrow A_1 + A_2$ does not transfer baryon number into the dark sector.  But because there are multiple possible initial states, the contribution from the decay of a particle can cancel with an interference term in 2-to-2 scattering.  The square of the bottom right diagram is removed by the subtraction of real intermediate states, but the interference term is not.
The CP asymmetry from the top two diagrams and the bottom two diagrams are
\bea
\dot{n}_{B,A_3 \rightarrow B_1 + B_2} &=& \int d\Pi\text{ } k \text{Im}(\lambda_1 \lambda_3^* g_1^*) f_{A_3} (1 \pm f_{A_1}) (1 \pm f_{A_2}) (1 \pm f_{B_1}) (1 \pm f_{B_2}) \\
\dot{n}_{B,A_1 + A_2 \rightarrow B_1 + B_2} &=& \int d\Pi \text{ } k \text{Im}(\lambda_1^* \lambda_3 g_1) f_{A_1} f_{A_2} (1 \pm f_{A_3}) (1 \pm f_{B_1}) (1 \pm f_{B_2})
\eea
where $k$ is a constant times momentum conserving delta functions.  In particular
\bea
\frac{\dot{n}_{B,A_3 \rightarrow B_1 + B_2}}{\dot{n}_{B,A_1 + A_2 \rightarrow B_1 + B_2}} = \frac{\text{Im}(\lambda_1 \lambda_3^* g_1^*)}{ \text{Im}((\lambda_1 \lambda_3^* g_1^*)^*)} \frac{\int d\Pi k f_{A_3} (1 \pm f_{A_1}) (1 \pm f_{A_2}) }{\int d\Pi k f_{A_1} f_{A_2} (1 \pm f_{A_3})}
\eea
Because there exists a process $\mathcal{M}(A_3 \rightarrow A_1 + A_2)$, thermal equilibrium of the visible sector enforces $f_{A_3} (1 \pm f_{A_1}) (1 \pm f_{A_2}) =  f_{A_1} f_{A_2} (1 \pm f_{A_3})$.  The two CP asymmetries cancel and there is no $\OO(\lambda^2)$ abundance generated.
As will be shown in Sec.~\ref{Sec: Unitarity} and again in Sec.~\ref{Sec: Diagrams}, these cancellations are not a coincidence and are required by CPT and unitarity.

While charges such as baryon number are not transferred at $\OO(\lambda^2)$, the CP violating interactions do generate asymmetries between particle series.  As shown in Fig.~\ref{Fig: BDD}, there will be equal and opposite CP violation for ($B_0$,$B_3$) and ($B_1$,$B_2$).  For the assumption of thermal equilibrium to hold as time proceeds means that CP symmetric collisions such as $g_2$ must wash out the asymmetry rapidly.  The $\OO(\lambda^2)$ deviation from thermal equilibrium can use CP symmetric interactions to transfer the symmetry back to the $A$ sector.  This effect is $\OO(\lambda^4)$.  This and other deviations from thermal equilibrium are discussed in detail in Sec.~\ref{Sec: Deviations}.

\section{Unitarity and CPT}
\label{Sec: Unitarity}

The cancellations illustrated in Sec.~\ref{Sec: Example} are a consequence of unitarity.  This section shows the implications of unitarity for the transfer of $U(1)$ charges between two sectors not in thermal equilibrium with each other. 
In the context of finite densities, unitarity is modified~\cite{Weinberg:1979bt}.  For states $\alpha$ and $\beta$, unitarity combined with CPT reads
\bea
& &\hspace{-0.5in}\sum_\beta \int d \Pi_\beta \delta^4(p_\alpha-p_\beta) |M(\alpha \rightarrow \beta)|^2 (1 \pm f_\beta) \CR
\label{Eq: decay}
& &\hspace{1in}= \sum_\beta \int d \Pi_\beta \delta^4(p_\alpha-p_\beta) |M(\bar{\alpha} \rightarrow \beta)|^2 (1 \pm f_\beta) \\
& &\hspace{-0.5in} \sum_\beta \int d \Pi_\beta \delta^4(p_\alpha-p_\beta) |M(\beta \rightarrow \bar{\alpha})|^2 (1 \pm f_\beta) \CR
 \label{Eq: decay2}
& &\hspace{1in}= \sum_\beta \int d \Pi_\beta \delta^4(p_\alpha-p_\beta) |M(\beta \rightarrow \alpha)|^2 (1 \pm f_\beta)
\eea
The stimulated emission/Pauli suppression factors were defined in Eq.~\ref{Eq: phase2}.  In the case of single particle states $\alpha$, Eq.~\ref{Eq: decay} states the familiar fact that particles and antiparticles have the same decay width.  Eq.~\ref{Eq: decay2} is the CPT conjugate of this fact with the sum over $\bar{\beta}$ replaced by the sum over $\beta$.  Assuming thermal equilibrium, we can use Eq.~\ref{Eq: expo} to replace $(1 \pm f_\beta)$ by $f_\beta$ on both sides of Eq.~\ref{Eq: decay2}.

As an illustrative warm-up, we first show that the transfer of conserved charges between two sectors in thermal equilibrium with each other is zero.  Let $Q$ denote the charge of a state.  The rate of change of $Q$ in the visible sector is
\bea
\label{Eq: Step1}
& &\hspace{-0.5in}\dot{n}_{Q} \propto   \sum_{\beta,\alpha} \int d\Pi_{\beta} d\Pi_{\alpha}  (2 \pi)^4 \delta^4(p_\alpha - p_\beta) \Delta |M(\alpha_v \alpha_h \rightarrow \beta_v \beta_h)|^2 \CR
& &\hspace{2.5in} \times f_{\alpha} (1 \pm f_\beta)  (Q(\beta_v)-Q(\alpha_v))
\eea 
The unitarity relation Eq.~\ref{Eq: decay} states
\bea
\sum_{\beta} \int d\Pi_{\beta} d\Pi_{\alpha} (2 \pi)^4 \delta^4(p_\alpha - p_\beta) \Delta |M(\alpha_v \alpha_h \rightarrow \beta_v \beta_h)|^2 f_{\alpha} (1 \pm f_\beta)  Q(\alpha_v)=0
\eea
Similarly Eq.~\ref{Eq: expo} and Eq.~\ref{Eq: decay2} require 
\bea
\sum_{\alpha} \int d\Pi_{\beta} d\Pi_{\alpha} (2 \pi)^4 \delta^4(p_\alpha - p_\beta) \Delta |M(\alpha_v \alpha_h \rightarrow \beta_v \beta_h)|^2 f_{\alpha} (1 \pm f_\beta) Q(\beta_v) = 0
\eea
Together these equations show that $\dot{n}_Q = 0$.  There is no net transfer of $U(1)$ charge between two sectors in thermal equilibrium with each other.  These cancellations by sums over final states or initial states in thermal equilibrium are seen in the diagrammatic approach to unitarity in Sec.~\ref{Sec: Diagrams} and was seen to a lesser extent in Sec.~\ref{Sec: Example}.

The situation described in this paper is that in which the two sectors are not in thermal equilibrium with each other.  Consider processes to $\OO(\lambda^2)$\footnote{For the decay of heavy particles, it was proven in \cite{Nanopoulos:1979gx} that baryon number generation in the freeze-out and decay scenario requires baryon number violating couplings to the third power.  This proof is a slight modification of that argument.}.  We have matrix elements 
\bea
\mathcal{M}(\alpha \rightarrow \beta) &=& M_{\beta \alpha} = \text{}_{\text{out}}\langle \beta_v \beta_h \mid H_\lambda \mid \alpha_v \alpha_h \rangle_{\text{in}} \\
\mathcal{M}(\alpha \rightarrow \beta)^* &=& M_{\beta \alpha}^*  = M_{\alpha \beta}^\dagger  = \text{}_{\text{in}}\langle \alpha_v \alpha_h \mid H_\lambda \mid \beta_v \beta_h \rangle_{\text{out}} \\
|\mathcal{M}(\alpha \rightarrow \beta)|^2 &=& M_{\beta \alpha} M_{\alpha \beta}^\dagger
\eea
 For the antiparticle process,  
\bea
\mathcal{M}(\bar{\alpha} \rightarrow \bar{\beta}) &=& \mathcal{M}(\beta \rightarrow \alpha)
= \text{}_{\text{out}}\langle \alpha_v \alpha_h \mid H_\lambda \mid \beta_v \beta_h \rangle_{\text{in}} \\
&=& \sum_{\delta,\gamma} \text{}_{\text{out}}\langle \alpha_v \alpha_h \mid \delta_v \delta_h \rangle_{\text{in}}  \text{   }_{\text{in}}\langle \delta_v \delta_h \mid H_\lambda \mid \gamma_v \gamma_h \rangle_{\text{out}}  \text{   }_{\text{out}}\langle \gamma_v \gamma_h \mid \beta_v \beta_h \rangle_{\text{in}} \\
\label{Eq: antiparticle}
&=& \sum_{\delta,\gamma} S_{\alpha_v \delta_v} S_{\alpha_h \delta_h} \mathcal{M}_{\delta \gamma}^\dagger   S_{\gamma_v \beta_v} S_{\gamma_h \beta_h}
\eea
where in the first step, CPT invariance was used and the second step involves the insertion of a complete set of in and out states.  When summing over states there is an implicit phase space integration.  $S$ is the subtracted S-matrix in the limit $\lambda \rightarrow 0$ using the notation 
\bea
S_{\beta \alpha} &=& \text{}_{\text{out}}\langle \beta \mid \alpha \rangle_{\text{in}} \\
S_{\beta \alpha}^\dagger &=& \text{}_{\text{in}}\langle \beta \mid \alpha \rangle_{\text{out}}
\eea
It satisfies the unitarity constraint\cite{Nanopoulos:1979gx}
\bea
\label{Eq: Unitarity1}
\sum_{\gamma_v} S_{\beta_v \gamma_v}^\dagger (1 \pm f_{\gamma_v}) S_{\gamma_v \alpha_v} = \delta_{\alpha_v \beta_v} (1 \pm f_{\alpha_v})
\eea
The phase space factor on the left side of the equation results from Pauli exclusion/stimulated emission, while the phase space factor on the right side of the equation results because a fermion is not restricted by the Pauli exclusion principle from freely propagating (trivial S matrix).  If the states involved are in thermal equilibrium then using Eq.~\ref{Eq: expo} and conservation laws, the unitarity relation can be rewritten as 
\bea
\label{Eq: Unitarity2}
\sum_{\gamma_v} S_{\beta_v \gamma_v}^\dagger f_{\gamma_v} S_{\gamma_v \alpha_v} = \delta_{\alpha_v \beta_v} f_{\alpha_v}
\eea

Consider a specific negative baryon number transferring channel ($Q(\alpha_v) - Q(\beta_v) = -B$).  Baryon number is a conserved quantity so that unitarity still holds when restricted to a specific baryon number state.  The baryon number transfer is proportional to
\bea
\nonumber
& &\hspace{-0.8in}\sum_{\alpha, \beta}  |\mathcal{M}(\bar{\alpha} \rightarrow \bar{\beta})|^2 f_{\alpha_v} f_{\alpha_h} (1 \pm f_{\beta_v}) (1 \pm f_{\beta_h}) \delta(Q(\alpha_v) - Q(\beta_v) = -B) \\
\label{Eq: initial}
&=&  \sum_{\alpha, \beta, \delta, \gamma, \rho, \phi} S_{\alpha_v \delta_v} S_{\alpha_h \delta_h} \mathcal{M}_{\delta \gamma}^\dagger   S_{\gamma_v \beta_v} S_{\gamma_h \beta_h}  \delta(Q(\delta_v) - Q(\gamma_v) = -B)\\
\nonumber
 &\times& S_{\rho_v \alpha_v}^\dagger S_{\rho_h \alpha_h}^\dagger \mathcal{M}_{\phi \rho}   S_{\beta_v \phi_v}^\dagger S_{\beta_h \phi_h}^\dagger f_{\alpha_v} f_{\alpha_h} (1 \pm f_{\beta_v}) (1 \pm f_{\beta_h}) 
\eea
In this first step, Eq.~\ref{Eq: antiparticle} is used.  It is critical that $S$ is the $\lambda \rightarrow 0$ limit of the S-matrix so that it treats the two sectors independently.  This fact allows us to use conservation of U(1) charge and relabel the charges so that unitarity relations, which involve sums over $\alpha$ and $\beta$, can be used.  Using the unitarity relation in Eq.~\ref{Eq: Unitarity1} simplifies the expression into
\bea
\nonumber
& &\hspace{-0.8in}\sum_{\alpha, \beta}  |\mathcal{M}(\bar{\alpha} \rightarrow \bar{\beta})|^2 f_{\alpha_v} f_{\alpha_h} (1 \pm f_{\beta_v}) (1 \pm f_{\beta_h}) \delta(Q(\alpha_v) - Q(\beta_v) = -B) \\
\label{Eq: initial2}
&=& \sum_{\alpha, \delta, \gamma, \rho, \phi} S_{\alpha_v \delta_v} S_{\alpha_h \delta_h} \mathcal{M}_{\delta \gamma}^\dagger \delta(Q(\delta_v) - Q(\gamma_v) = -B) \\
\nonumber
&\times& S_{\rho_v \alpha_v}^\dagger S_{\rho_h \alpha_h}^\dagger \mathcal{M}_{\gamma \rho}  f_{\alpha_v} f_{\alpha_h} (1 \pm f_{\gamma_v}) (1 \pm f_{\gamma_h})  \delta_{\phi \gamma}
\eea
Finally using Eq.~\ref{Eq: Unitarity2} gives the result
\bea
\nonumber
& &\hspace{-0.8in}\sum_{\alpha, \beta}  |\mathcal{M}(\bar{\alpha} \rightarrow \bar{\beta})|^2 f_{\alpha_v} f_{\alpha_h} (1 \pm f_{\beta_v}) (1 \pm f_{\beta_h}) \delta(Q(\alpha_v) - Q(\beta_v) = -B) \\
\label{Eq: initial3}
\hspace{0.2in} &=&   \sum_{\delta, \gamma} |\mathcal{M}(\delta \rightarrow \gamma)|^2  f_{\delta_v} f_{\delta_h} (1 \pm f_{\gamma_v}) (1 \pm f_{\gamma_h}) \delta(Q(\delta_v) - Q(\gamma_v) = -B) \\
\hspace{0.2in} &=&  \sum_{\alpha, \beta}  |\mathcal{M}(\alpha \rightarrow \beta)|^2 f_{\alpha_v} f_{\alpha_h} (1 \pm f_{\beta_v}) (1 \pm f_{\beta_h}) \delta(Q(\alpha_v) - Q(\beta_v) = -B) 
\eea
The baryon number transfer is equal and opposite to the baryon number transfer of the $+B$ sector, $\sum \Delta \mathcal{M} = 0$.  Thus there is no baryon number transferred between the two sectors at $\OO(\lambda^2)$.

\section{Feynman Diagrams}
\label{Sec: Diagrams}

In Sec.~\ref{Sec: Unitarity}, it was proven that the baryon number generation vanishes to leading order.  The same results can be seen diagrammatically.  The diagrammatic approach, while more involved, lets one show exactly how an asymmetry is generated and places additional constraints on the two sectors.

For simplicity, we will assume that the hidden sector is at zero temperature.
As which states are in the Cutkosky cuts is very important, I will change notation slightly.  Before, the general state was $\alpha = \alpha_v \otimes \alpha_h$.  Rather than explicitly writing out $\alpha_v \otimes \Omega_h$, where $\Omega_h$ is the vacuum state for the hidden sector, I will simply represent it as $\alpha_v$.  The state $\alpha_v \alpha_h$ will only be written when both $\alpha_v \ne \Omega_v$ and $\alpha_h \ne \Omega_h$.

The U(1) charge transfer between the two sectors is
\bea
& &\hspace{-0.5in}\dot{n}_B \propto   \sum_{\alpha_v, \beta} \int d\Pi_{\alpha_v} d\Pi_{\beta} (2 \pi)^4 \delta^4(p_{\alpha_v} - p_{\beta}) \Delta  |\mathcal{M}(\alpha_v \rightarrow \beta)|^2 \CR
& &\hspace{2.5in} \times f_{\alpha_v} (1 \pm f_{\beta}) (Q(\beta_v)-Q(\alpha_v)) 
\eea
Diagrams can be grouped by their Cutkosky cuts\footnote{For the sake of simplicity, only $s$ channel cuts are considered.}.  There are six different types of interference terms depending on if the intermediate state $\delta$ is $\delta_v$,$\delta_h$ or $\delta_v \delta_h$ and whether the final state is $\beta_h$ or $\beta_v \beta_h$.

\begin{figure}[h] 
   \centering
   \includegraphics[width=4in]{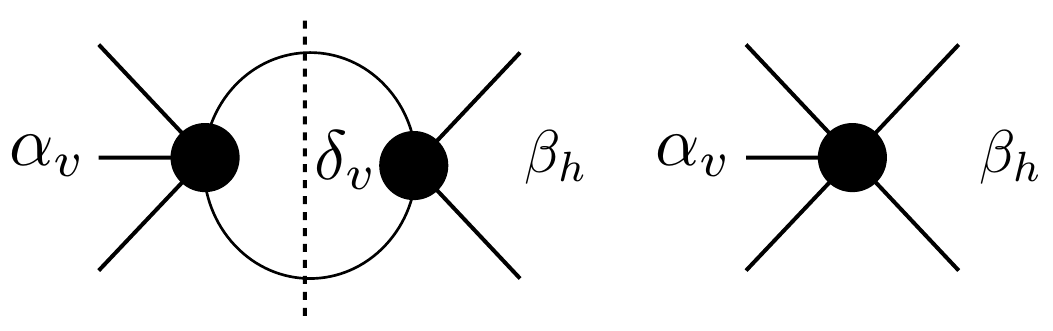} 
   \caption{The CP violation comes from this interference where the $\delta_v$ intermediate state can go onshell.}
   \label{Fig: BB'D}
\end{figure}

\begin{figure}[h] 
   \centering
   \includegraphics[width=4in]{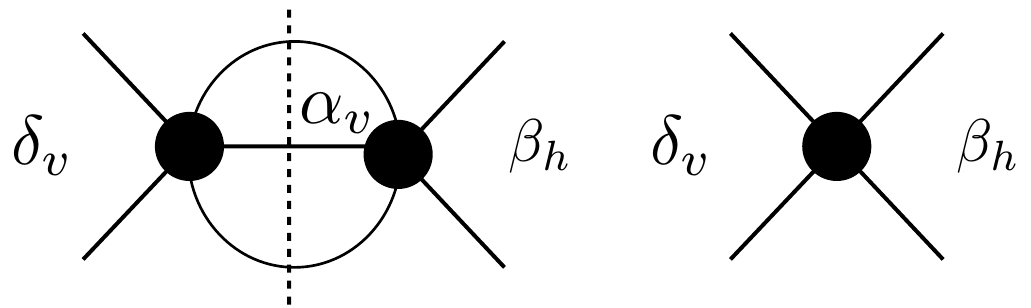} 
   \caption{Baryon asymmetry generated from this diagram cancels the baryon asymmetry generated from the figure \ref{Fig: BB'D} }
   \label{Fig: B'BD}
\end{figure}

Consider the $\OO(\lambda^2)$ case where the intermediate particle is a state $\delta_v$ and the final state is either $\beta_h$ or $\beta_v \beta_h$.  For concreteness, consider the final state $\beta_h$.  Using the results of Eq.~\ref{Eq: interference} gives
\bea
\dot{n}_B &\propto& \sum_{\alpha_v,\delta_v} Q(\alpha_v) f_{\alpha_v} (1 \pm f_{\beta_h}) (1 \pm f_{\delta_v}) \text{Im}[ \lambda_{\alpha_v \rightarrow \beta_h} g_{\alpha_v \rightarrow \delta_v}^* \lambda_{\delta_v \rightarrow \beta_h}^* ] F_{\alpha_v \rightarrow \beta_h} F_{\alpha_v \rightarrow \delta_v} F_{\delta_v \rightarrow \beta_h} 
\eea
Only the relevant terms are included in the sum and everything else was pushed into the proportionality constant.  The sum can be simplified by realizing that the kinematic part of the Feynman diagrams, the phase space factors and the charge are all symmetric under $\alpha_v \leftrightarrow \delta_v$.  Conservation of charge allows the swap to have no effect on the charge while the phase space factors obey $f_{\alpha_v} (1 \pm f_{\delta_v}) = f_{\delta_v} (1 \pm f_{\alpha_v}) $ due to the assumption of equilibrium values.  Kinematics are independent of which state is the initial and which is the final so the $F$ are invariant.  On the other hand, the couplings are odd under $\alpha_v \leftrightarrow \delta_v$.  Relabeling the states changes the couplings by  
\bea
\text{Im}[ \lambda_{\alpha_v \rightarrow \beta_h} g_{\alpha_v \rightarrow \delta_v}^* \lambda_{\delta_v \rightarrow \beta_h}^*]  + \text{Im}[  \lambda_{\delta_v \rightarrow \beta_h} g_{\delta_v \rightarrow \alpha_v}^* \lambda_{\alpha_v \rightarrow \beta_h}^* ] = 0
\eea
Since the coupling constants were odd under this transformation while the rest of the factors were invariant, the sum vanishes.  This cancellation was seen earlier in Sec.~\ref{Sec: Unitarity} as a sum over initial states combined with thermal equilibrium.
It is critical for thermal field theory to be used in the calculation in order to see the cancellation.  This cancellation occurs diagram by diagram with each pair being obtained just by flipping the intermediate state and the initial state.
This cancellation is shown graphically in Figs. \ref{Fig: BB'D} and \ref{Fig: B'BD}

%

A similar cancellation occurs for the other $\OO(\lambda^2)$ case where the final state is a state $\beta_h$ and the intermediate state is a state $\delta_h$.
\bea
\dot{n}_B &\propto& \sum_{\beta_h,\delta_h} (1 \pm f_{\beta_h})(1 \pm f_{\delta_h}) \text{Im}[ \lambda_{\alpha_v \rightarrow \beta_h} \lambda_{\alpha_v \rightarrow \delta_h}^* g_{\delta_h \rightarrow \beta_h}^* ]  F_{\alpha_v \rightarrow \beta_h} F_{\alpha_v \rightarrow \delta_h} F_{\delta_h \rightarrow \beta_h} \\ 
&=& 0
\eea
As before, this cancellation relies on the fact that the couplings are odd under $\beta_h \leftrightarrow \delta_h$ while the rest of the factors are even.  This cancellation was seen in Sec.~\ref{Sec: Unitarity} as a sum over final states.  Again the cancellation is in pairs of diagrams obtained by flipping the intermediate and final states.  These two cases are the only $\OO(\lambda^2)$ contributions to the baryon asymmetry and they both cancel completely.

Things are different at $\OO(\lambda^3)$ where the intermediate state is $\delta_v \delta_h$.  By the previous argument, one might expect that the CP violation from $\alpha_v \rightarrow \delta_v \delta_h \rightarrow \beta_h$ would cancel with $\alpha_v \rightarrow \beta_h \rightarrow \delta_v \delta_h$.  This cancellation would occur except for the fact that final states $\delta_v \delta_h$ and $\beta_h$ transfer differing amounts of baryon number to the hidden sector.  More explicitly, the sum of $\alpha_v \rightarrow \delta_v \delta_h \rightarrow \beta_h$ and $\alpha_v \rightarrow \beta_h \rightarrow \delta_v \delta_h$ goes as 
\bea
\dot{n}_B &\propto& \sum_{\beta_h,\delta_v \delta_h} Q(\alpha_v) (1 \pm f_{\beta_h})(1 \pm f_{\delta_v \delta_h}) \text{Im}[ \lambda_{\alpha_v \rightarrow \beta_h} \lambda_{\alpha_v \rightarrow \delta_v \delta_h}^* \lambda_{\delta_v \delta_h \rightarrow \beta_h}^* ] \CR
& & \hspace{3in} \times F_{\alpha_v \rightarrow \beta_h}  F_{\alpha_v \rightarrow \delta_v \delta_h} F_{\delta_v \delta_h \rightarrow \beta_h} \\ 
\nonumber
&+& \sum_{\beta_h,\delta_v \delta_h} (Q(\alpha_v)-Q(\delta_v)) (1 \pm f_{\beta_h})(1 \pm f_{\delta_v \delta_h})\text{Im}[ \lambda_{\alpha_v \rightarrow \delta_v \delta_h} \lambda_{\alpha_v \rightarrow \beta_h}^* \lambda_{\beta_h \rightarrow \delta_v \delta_h}^* ] \CR
& & \hspace{3in} \times F_{\alpha_v \rightarrow \delta_v \delta_h} F_{\alpha_v \rightarrow \beta_h} F_{\beta_h \rightarrow \delta_v \delta_h} \\
&\propto& \sum_{\beta_h,\delta_v \delta_h} Q(\delta_v) (1 \pm f_{\beta_h})(1 \pm f_{\delta_v \delta_h})\text{Im}[ \lambda_{\alpha_v \rightarrow \beta_h} \lambda_{\alpha_v \rightarrow \delta_v \delta_h}^* \lambda_{\delta_v \delta_h \rightarrow \beta_h}^* ] \CR
& & \hspace{3in} \times F_{\alpha_v \rightarrow \beta_h} F_{\alpha_v \rightarrow \delta_v \delta_h} F_{\delta_v \delta_h \rightarrow \beta_h}
\eea
and so a contribution is left that does not cancel.  Similarly, for a state where the imaginary component comes from $\alpha_v \rightarrow \delta_v \delta_h \rightarrow \beta_v \beta_h$, the result is the same where $\beta_h$ is replaced by $\beta_v \beta_h$ and is weighted by $Q(\delta_v)-Q(\beta_v)$.  If the two sectors were in thermal equilibrium with each other, these CP violation effects would be cancelled in a sum over initial states, like the first case considered above.  

All of the $\OO(\lambda^2)$ effects have cancelled and an asymmetry is generated at $\lambda^3$.    In minimal cases $\lambda$ itself must be CP violating in order for this reaction to cause a CP asymmetry.  This result is similar in spirit to those in Ref. \cite{Nanopoulos:1979gx} where it was shown that baryon number violation effects come in at cubic order in baryon number violating couplings.

\section{Deviations from thermal equilibrium}
\label{Sec: Deviations}

One has to be careful that there are not any $\OO(\lambda^2)$ higher order effects which have not been cancelled.  The previous computations showed that whenever the phase space factors are not modified, the baryon number generated is $\OO(\lambda^3)$.  There are multiple ways to change the phase space factors.  One way is to include the effect of $\lambda$ on the phase space factors.  The phase space factors were assumed to be thermally distributed with different temperatures for the different sectors.  Because there exist matrix elements connecting the two sectors, the phase space factors are modified to $f_{\alpha_v} = f_{\alpha_v}^{\text{eq}} + \delta f_{\alpha_v}$.  The deviation $\delta f_{\alpha_v} \propto |\mathcal{M}(\alpha_v \rightarrow \beta_v \beta_h)|^2 \propto \lambda^2$.  The baryon asymmetry generated by this deviation is $\OO(\lambda^4)$ and is therefore subdominant.  

Another possible effect comes from the fact that $f_{\alpha}$ will leave thermal equilibrium due to Hubble expansion; particles can freeze-out.  The baryon asymmetries generated from freeze-out typically scale as $\OO(\lambda^2)$ but with non-negligible coefficients.  Freeze-out abundances scale differently than freeze-in abundances so which effect dominates is a model dependent numerical question.

A case where freeze-out asymmetries always dominate over freeze-in asymmetries is a visible sector hot relic which becomes stable as $\lambda$ goes to zero.  Because the width goes to zero as $\lambda$ goes to zero, the intermediate on-shell states must contain both visible and hidden sector particles (hidden sector only intermediate states would have canceling baryon asymmetries when summed over all possible final states).  The decay width is $\OO(\lambda^2)$ while the CP violating decay widths are $\OO(\lambda^3)$, thus the CP violating branching ratio into the dark sector scales as $\lambda$ rather than $\lambda^2$.  The asymmetry that results from freeze-out and decay would be at best the same order effect as the freeze-in asymmetry and in most cases dominate the freeze-in asymmetry.  
Other relics typically have $\OO(\lambda^2)$ CP violating branching ratios into the dark sector and the issue which mechanism dominates is completely numerical.
In models of freeze-in baryogenesis, freeze-out typically yields CP asymmetries orders of magnitude below the freeze-in asymmetries.

The cross section for production of particles in the dark sector is $\OO(\lambda^2)$ while the CP asymmetry is $\OO(\lambda^3)$.  Not overclosing the universe imposes strict constraints. The dark matter particles would need to be very light (eV or less) if there is an $\OO(\lambda^2)$ abundance.  If there is an efficient annihilation mechanism removing the $\OO(\lambda^2)$ component~\cite{Graesser:2011wi}, the dark matter is MeV scale or less and the model joins the long history of asymmetric dark matter~\cite{Nussinov:1985xr,Kaplan:1991ah,Kaplan:2009ag,Falkowski:2011xh}.
Being able to annihilate well imposes constraints, for example, the dark sector must not have any stable heavy particles that do not carry baryon number.  

\begin{figure}[h] 
   \centering
   \includegraphics[width=4in]{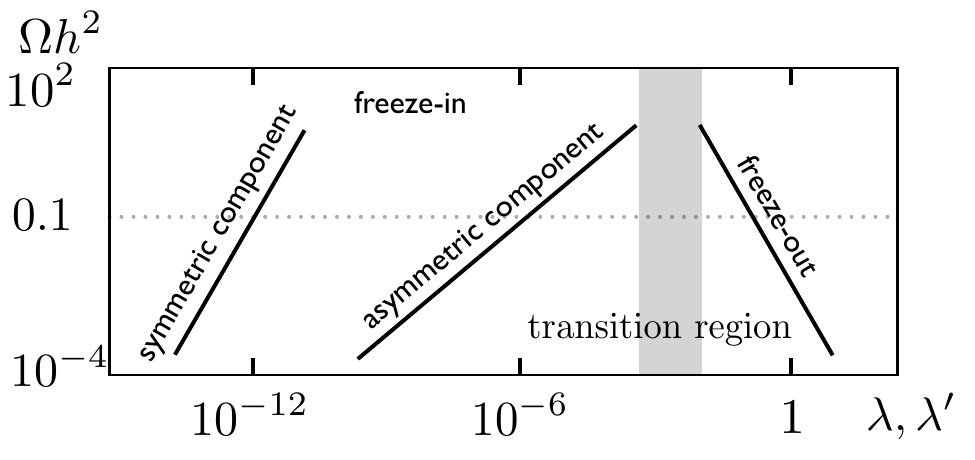} 
   \caption{A schematic drawing of how the different abundances scale.  The asymmetric abundance generated by freeze-in lags behind the symmetric component by a factor of $\lambda$.}
   \label{Fig: Tree}
\end{figure}

\section{Conclusion}

This note has shown that if two sectors are in thermal equilibrium with themselves, but not with each other, then the leading effect transferring conserved quantities between the two sectors is of order the the weak coupling connecting them to the third power.  When freeze-in is used to produce a net baryon number density, the leading order effect comes from $\OO(\lambda^3)$ diagrams where the intermediate state that goes on-shell has a different visible baryon number than the final state visible baryon number.  Models in which the correct baryon number is generated with freeze-in as the dominant source of abundance, typically require $\lambda \gsim 10^{-6}$ and $m_{\text{bath}} \gsim$ TeV.  $m_{\text{bath}}$ is the mass of the visible particle which communicates with the hidden sector.  The lower window is potentially observable at the LHC.

\section*{Acknowledgments} 
AH would like to thank Lawrence Hall, John March-Russell, and Stephen West for helpful discussions and Shamit Kachru for comments on the draft.  AH would like to thank Jay Wacker for a critical reading of the draft and especially thanks Michael Peskin for both very helpful discussions and plentiful comments on the draft.  The work was supported by the US Department of Energy under contract DE-AC02-76SF00515.

\appendix
\section{CP violation by diagrams regulated by the width}
\label{Sec: B}

In Sec.~\ref{Sec: sub}, it was claimed that higher order diagrams could cancel lower order diagrams for unsubtracted matrix elements.  In the case of nonsubtracted matrix elements, the diagrams which cancel are shown in Fig. \ref{Fig: Tree}.  A loop induced CP violation cancels tree level CP violation with the narrow width approximation off setting the use of a loop.  The width is an infinite sum of diagrams.  In this manner, the CP violation from an infinite set of diagrams cancel the CP violation of a single tree level diagram.


\begin{figure}[h] 
   \centering
   \includegraphics[width=4in]{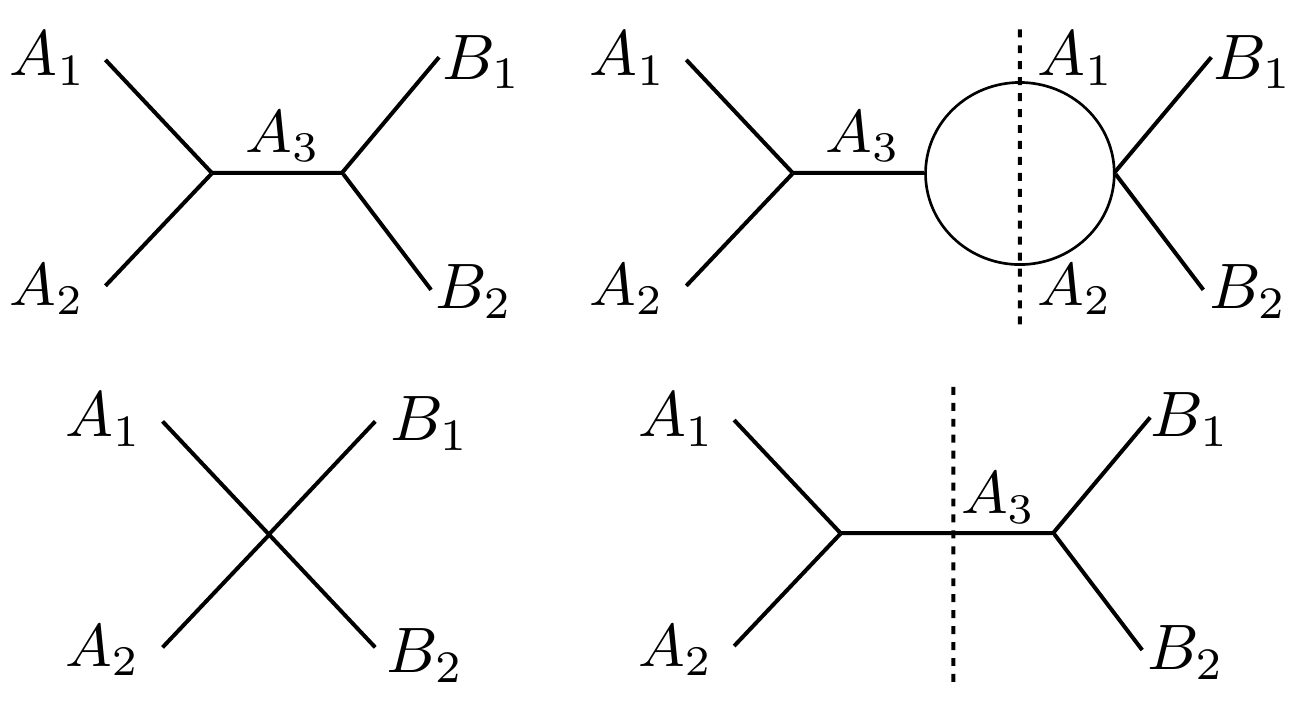} 
   \caption{When the top diagram is resonantly produced and the narrow width approximation is used, a factor of $\frac{1}{\Gamma} \sim |\mathcal{M}(A_3 \rightarrow A_1 + A_2)|^{-2}$ appears.  This inverse coupling allows the higher order diagram to cancel against a single tree level diagram.  The infinite number of diagrams that go into a width are responsible for canceling the tree level CP violation.  Subtraction of real intermediate states simplifies the cancellation of CP asymmetries.}
   \label{Fig: Tree}
\end{figure}

%

\end{document}

%% file: mymacros.tex
\def\beq{\begin{equation}}
\def\eeq#1{\label{#1}\end{equation}}
\def\eeqn{\end{equation}}

\newenvironment{Eqnarray}%
   {\arraycolsep 0.14em\begin{eqnarray}}{\end{eqnarray}}
\def\beqa{\begin{Eqnarray}}
\def\eeqa#1{\label{#1}\end{Eqnarray}}
\def\eeqan{\end{Eqnarray}}
\def\CR{\nonumber \\ }

\let\bar=\overbar

\def\lsim{\mathrel{\raise.3ex\hbox{$<$\kern-.75em\lower1ex\hbox{$\sim$}}}}
\def\gsim{\mathrel{\raise.3ex\hbox{$>$\kern-.75em\lower1ex\hbox{$\sim$}}}}

\def\del{\partial}
\def\Dslash{\not{\hbox{\kern-4pt $D$}}}
\def\dslash{\not{\hbox{\kern-2pt $\del$}}}

\def\msb{{\bar{\scriptsize M \kern -1pt S}}}

\def\drb{{\bar{\scriptsize D \kern -1pt R}}}

\makeatletter
\def\section{\@startsection{section}{0}{\z@}{5.5ex plus .5ex minus
 1.5ex}{2.3ex plus .2ex}{\large\bf}}
\def\subsection{\@startsection{subsection}{1}{\z@}{3.5ex plus .5ex minus
 1.5ex}{1.3ex plus .2ex}{\normalsize\bf}}
\def\subsubsection{\@startsection{subsubsection}{2}{\z@}{-3.5ex plus
-1ex minus  -.2ex}{2.3ex plus .2ex}{\normalsize\sl}}

\renewcommand{\@makecaption}[2]{%
   \vskip 10pt
   \setbox\@tempboxa\hbox{\small #1: #2}
   \ifdim \wd\@tempboxa >\hsize     
       \small #1: #2\par          
     \else                        
       \hbox to\hsize{\hfil\box\@tempboxa\hfil}
   \fi}

 \def\citenum#1{{\def\@cite##1##2{##1}\cite{#1}}}
 
\newcount\@tempcntc
\def\@citex[#1]#2{\if@filesw\immediate\write\@auxout{\string\citation{#2}}\fi
  \@tempcnta\z@\@tempcntb\m@ne\def\@citea{}\@cite{\@for\@citeb:=#2\do
    {\@ifundefined
       {b@\@citeb}{\@citeo\@tempcntb\m@ne\@citea\def\@citea{,}{\bf ?}\@warning
       {Citation `\@citeb' on page \thepage \space undefined}}%
    {\setbox\z@\hbox{\global\@tempcntc0\csname b@\@citeb\endcsname\relax}%
     \ifnum\@tempcntc=\z@ \@citeo\@tempcntb\m@ne
       \@citea\def\@citea{,}\hbox{\csname b@\@citeb\endcsname}%
     \else
      \advance\@tempcntb\@ne
      \ifnum\@tempcntb=\@tempcntc
      \else\advance\@tempcntb\m@ne\@citeo
      \@tempcnta\@tempcntc\@tempcntb\@tempcntc\fi\fi}}\@citeo}{#1}}
\def\@citeo{\ifnum\@tempcnta>\@tempcntb\else\@citea\def\@citea{,}%
  \ifnum\@tempcnta=\@tempcntb\the\@tempcnta\else
  {\advance\@tempcnta\@ne\ifnum\@tempcnta=\@tempcntb \else\def\@citea{--}\fi
    \advance\@tempcnta\m@ne\the\@tempcnta\@citea\the\@tempcntb}\fi\fi}
\makeatother